\documentclass[article,nojss,shortnames]{jss}

\usepackage{thumbpdf}
\usepackage{amsfonts,amstext,amsmath,amssymb,amsthm}
\usepackage{accents}
\usepackage{color}
\usepackage{rotating}
\usepackage{verbatim}
\usepackage{rotfloat}
\usepackage{booktabs}
\usepackage{multirow}

\usepackage{caption}
\usepackage{longtable,pdflscape}

\newenvironment{knitrout}{}{} 
\definecolor{fgcolor}{rgb}{0.345, 0.345, 0.345}

\renewcommand{\Prob}{\mathbb{P}}

\usepackage{booktabs}

\newcommand{\rZ}{\text{treatment}}
\newcommand{\rY}{T}
\newcommand{\rX}{\mX}
\newcommand{\rz}{r}
\newcommand{\ry}{t}
\newcommand{\rx}{\xvec}

\newcommand{\samX}{\chi}

\newcommand{\s}{\svec}

\newcommand{\basisy}{\avec}
\newcommand{\bern}[1]{\avec_{\text{Bs}}}

\newcommand{\parm}{\varthetavec}
\newcommand{\eparm}{\vartheta}
\newcommand{\dimparm}{P}

\newcommand{\shiftparm}{\betavec}

\newcommand{\eshiftparm}{\beta}

\newcommand{\ie}{\textit{i.e.}~}
\newcommand{\eg}{\textit{e.g.}~}

\newcommand{\RR}{\mathbb{R}}

\usepackage{dsfont}
\newcommand{\I}{\mathds{1}}

 \DeclareMathOperator*{\argmax}{{arg\,max}}

 \DeclareMathOperator{\WD}{W}

\def \avec {\text{\boldmath$a$}}

\def \svec {\text{\boldmath$s$}}

\def \xvec {\text{\boldmath$x$}}    \def \mX {\text{\boldmath$X$}}

\def \alphavec        {\text{\boldmath$\alpha$}}
\def \betavec         {\text{\boldmath$\beta$}}

\def \varthetavec     {\text{\boldmath$\vartheta$}}

\newcommand{\ubar}[1]{\underaccent{\bar}{#1}}

\usepackage{algorithm}

\usepackage[]{graphicx}\usepackage[]{color}
\makeatletter
\def\maxwidth{ %
  \ifdim\Gin@nat@width>\linewidth
    \linewidth
  \else
    \Gin@nat@width
  \fi
}
\makeatother

\renewcommand{\maxwidth}{.8\textwidth}

\definecolor{fgcolor}{rgb}{0.345, 0.345, 0.345}
\newcommand{\hlnum}[1]{\textcolor[rgb]{0.686,0.059,0.569}{#1}}%
\newcommand{\hlstr}[1]{\textcolor[rgb]{0.192,0.494,0.8}{#1}}%
\newcommand{\hlcom}[1]{\textcolor[rgb]{0.678,0.584,0.686}{\textit{#1}}}%
\newcommand{\hlopt}[1]{\textcolor[rgb]{0,0,0}{#1}}%
\newcommand{\hlstd}[1]{\textcolor[rgb]{0.345,0.345,0.345}{#1}}%
\newcommand{\hlkwb}[1]{\textcolor[rgb]{0.69,0.353,0.396}{#1}}%
\newcommand{\hlkwc}[1]{\textcolor[rgb]{0.333,0.667,0.333}{#1}}%
\newcommand{\hlkwd}[1]{\textcolor[rgb]{0.737,0.353,0.396}{\textbf{#1}}}%

\usepackage{framed}
\makeatletter
\newenvironment{kframe}{%
 \def\at@end@of@kframe{}%
 \ifinner\ifhmode%
  \def\at@end@of@kframe{\end{minipage}}%
  \begin{minipage}{\columnwidth}%
 \fi\fi%
 \def\FrameCommand##1{\hskip\@totalleftmargin \hskip-\fboxsep
 \colorbox{shadecolor}{##1}\hskip-\fboxsep
     \hskip-\linewidth \hskip-\@totalleftmargin \hskip\columnwidth}%
 \MakeFramed {\advance\hsize-\width
   \@totalleftmargin\z@ \linewidth\hsize
   \@setminipage}}%
 {\par\unskip\endMakeFramed%
 \at@end@of@kframe}
\makeatother

\definecolor{shadecolor}{rgb}{.97, .97, .97}
\definecolor{messagecolor}{rgb}{0, 0, 0}
\definecolor{warningcolor}{rgb}{1, 0, 1}
\definecolor{errorcolor}{rgb}{1, 0, 0}

\usepackage{alltt}
\IfFileExists{upquote.sty}{\usepackage{upquote}}{}
\usepackage{paralist}

\setkeys{Gin}{width=0.5\textwidth, keepaspectratio = TRUE}

\author{Natalia Korepanova \\ National Research University \\ Higher School of Economics, Moscow
  \And Heidi Seibold \\ Ludwig-Maximilians-Universit\"at \\ M\"unchen
  \AND Verena Steffen \\ Genentech, a Member of the Roche Group, \\ South San Francisco
  \And Torsten Hothorn \\ Universit\"at Z\"urich}
\Plainauthor{Korepanova, Seibold, Steffen \& Hothorn}

\title{Survival Forests under Test: \\ Impact of the Proportional Hazards Assumption on
       Prognostic and Predictive Forests for ALS Survival}
\Plaintitle{Survival Forests under Test}
\Shorttitle{Survival Forests under Test}

\Abstract{

We investigate the effect of the proportional hazards assumption on
prognostic and predictive models of the survival time of patients suffering
from amyotrophic lateral sclerosis (ALS).  We theoretically compare the
underlying model formulations of several variants of survival forests and
implementations thereof, including random forests for survival, conditional
inference forests, Ranger, and survival forests with $L_1$ splitting, with
two novel variants, namely distributional and transformation survival
forests.  Theoretical considerations explain the low power of log-rank-based
splitting in detecting patterns in non-proportional hazards situations in
survival trees and corresponding forests.  This limitation can potentially
be overcome by the alternative split procedures suggested herein.  We
empirically investigated this effect using simulation experiments and a
re-analysis of the PRO-ACT database of ALS survival, giving special emphasis
to both prognostic and predictive models.

}

\Keywords{transformation model, conditional survivor function, conditional hazard
function, survival trees, amyotrophic lateral sclerosis}
\Plainkeywords{transformation model, conditional survivor function, conditional hazard
function, survival trees, amyotrophic lateral sclerosis}

\Address{
  
  Natalia Korepanova \\
  International Laboratory for Intelligent Systems and Structural Analysis \\
  Faculty of Computer Science \\
  National Research University Higher School of Economics \\
  Kochnovsky Proezd 3, RU-427553 Moscow, Russia \\ 

  Heidi Seibold \\
  Institut f\"ur Medizinische Informationsverarbeitung, Biometrie und Epidemiologie \\
  Ludwig-Maximilians-Universit\"at M\"unchen \\
  Marchioninistra{\ss}e 15, DE-81377 M\"unchen, Germany \\

  Verena Steffen \\
  Genentech, Inc., 1 DNA Way, South San Francisco, CA 94080, U.S.A. \\

  Torsten Hothorn\\
  Institut f\"ur Epidemiologie, Biostatistik und Pr\"avention \\
  Universit\"at Z\"urich \\
  Hirschengraben 84, CH-8001 Z\"urich, Switzerland \\
  \texttt{Torsten.Hothorn@uzh.ch}
}

\begin{document}

\section{Introduction}

Amyotrophic lateral sclerosis (ALS) is a devastating neurodegenerative
disease.  The disease often progresses rapidly and leads to early death for
many patients.  The identification of prognostic factors and the subsequent
development of prognostic models forecasting disease progression have long
been difficult and vital problems.  The availability of such instruments
would, for example, allow the planning of more powerful clinical trials by
means of efficient patient stratification
\citep{Chio_Logroscino_Hardiman_2009}.  Two approaches have been used in the
past, namely the search for prognostic models for the overall survival time
after diagnosis \citep[][among many others]{Kimura_Fujimura_Ishida_2006,
Zoccolella_Beghi_Palagano_2008,FujimuraKiyono_Kimura_Ishida_2011,
Beaulieu_Green_2016,Mandrioli_Rosi_Fini_2017,Ong_Tan_Holbrook_2017,Pfohl_Kim_Coan_2018}
and the prognosis of a functional assessment of patients via the ordinal ALS
functional rating scale \citep[ALSFRS;][]{Brooks_Sanjak_Ringel_1996} and
ALSFRS-R scores
\citep{Cedarbaum_Stambler_Malta_1999,Hothorn_Jung_2014,Kueffner_Zach_Norel_2014}.

Riluzole (Rilutek) is the only approved drug for ALS treatment and
potentially prolongs median survival by a few months.  Predictive models,
\ie models describing the differential treatment effect of Riluzole as a
function of patient characteristics, are important for a better
understanding of the mechanisms of Riluzole interaction with the nervous
system.  To date, differential treatment effects of Riluzole have been
reported in traditional subgroup analyses
\citep{Fang_Khleifat_Meurgey_2018}, statistical learning approaches for
``automated'' subgroup analysis \citep{Seibold_Zeileis_Hothorn_2015}, and
estimation of individualized treatment effects
\citep{Seibold_Zeileis_Hothorn_2017}.

Random forests play an important role in these developments as many
researchers have applied variants of this method for building prognostic
\citep{Hothorn_Jung_2014,Beaulieu_Green_2016,Ong_Tan_Holbrook_2017,Pfohl_Kim_Coan_2018}
and predictive models \citep{Seibold_Zeileis_Hothorn_2017}.  It seems to be
a common belief that survival forest implementations such as random forest
for survival \citep[RF-S,][]{Ishwaran_Kogalur_Blackstone_2008}, conditional
inference forests \citep[CForest,][]{Hothorn_Lausen_Benner_2004,Hothorn_Zeileis_2015}, and
Ranger \citep{Wright_Ziegler_2017} ``handle the proportionality assumption
coherently and automatically'' \citep{Datema_Moya_Krause_2012}.  Similar
statements can be found in virtually every publication advocating the use of
survival forests over the application of traditional Cox proportional
hazards modelling.

However, a novel theoretical understanding of random forests as adaptive
local maximum-likelihood estimators
\citep{Athey_Tibshirani_Wager_2018,Hothorn_Zeileis_2017,Schlosser_Hothorn_Stauffer_2018}
highlights that this belief is overoptimistic.  In a nutshell, the log-rank
splitting \citep[as introduced by][for survival trees]{Segal_1988} typically
applied in survival forests
\citep{Hothorn_Lausen_Benner_2004,Ishwaran_Kogalur_Blackstone_2008,Wright_Dankowski_Ziegler_2017}
poses a challenge for survival trees: the detection of prognostic effects
whose impact on the conditional survivor function is not well described by a
shift on the log-cumulative hazards scale is difficult.  Consequently,
application of survival forests still requires careful assessment of the
impact of the proportional hazards assumption.  Here, we investigate the
impact of potential non-proportional hazards on prognostic and predictive
survival forest models of ALS.  We report on the performance of prognostic
and predictive survival models obtained under the classical log-rank
splitting as well as on the performance of three alternative survival forest
algorithms, all of which explicitly target the situation of
non-proportional hazards.

We theoretically deconstruct the myth that the proportional hazards
assumption is not an issue in survival forests in Section~\ref{sec:methods}. 
Using the flexible transformation forests framework
\citep{Hothorn_Zeileis_2017}, we design novel split criteria for prognostic
and predictive survival trees, which are powerful in both the proportional
and the non-proportional hazards setting.  We compared several variants of
survival forests in this class to established survival forests (RF-S,
CForest, Ranger) and to one recent proposal \citep[explicitly targeting the
non-proportional hazards situation using $L_1$
splitting,][]{Moradian_Larocque_Bellavance_2016} in an artificial prognostic
model setting and then investigated the empirical performance for ALS
survival prognosis.  In a second step, we compared predictive survival
forest models based on Weibull models \citep[``distributional survival
forests'', DSF,][]{Seibold_Zeileis_Hothorn_2017} with a less restrictive
novel variant of prognostic and predictive survival forests introduced
herein (``transformation survival forests'', TSF), both with respect to
predictive ALS models and based on simulations.

\section{Methods} \label{sec:methods}

A prognostic model for a survival time $\rY \in [0, \infty)$ describes the
impact of prognostic variables $\rX \in \samX$ available at time $\ry = 0$
on the conditional survivor function $1 - \Prob(\rY \le \ry \mid \rX =
\rx)$.  Without loss of generality, we can parameterize such a model as
\begin{eqnarray*}
\Prob(\rY \le \ry \mid \rX = \rx) = 
  1 - \exp\left(-\exp\left(\basisy(\ry)^\top \parm(\rx)\right)\right),
\end{eqnarray*}
where the log-cumulative hazard function $\basisy(\ry)^\top \parm(\rx)$ is
defined by some basis functions $\basisy: \RR^+ \rightarrow \RR^\dimparm$
and a conditional parameter function $\parm: \samX \rightarrow
\RR^\dimparm$.  The latter function is typically estimated based on data
from $i = 1, \dots, N$ independent subjects with prognostic variables $\rx_i
\in \samX$ and either an exact survival time $\rY = \ry_i \in [0, \infty)$,
a right-censored survival time $\rY > \ubar{\ry}_i \in [0, \infty)$, a
left-censored survival time $\rY < \bar{\ry}_i \in [0, \infty)$, or an
interval-censored survival time $\rY \in (\ubar{\ry}_i, \bar{\ry}_i] \subset
[0, \infty)$ under random censoring and possibly under some form of
truncation.

Random forests can be understood as local adaptive likelihood estimators for
the conditional parameter function $\parm(\rx)$ for a patient with
prognostic or predictive variables $\rx$
\citep{Athey_Tibshirani_Wager_2018,Hothorn_Zeileis_2017,Schlosser_Hothorn_Stauffer_2018}:
\begin{eqnarray} \label{fm:hatparm}
\hat{\parm}_N(\rx) = \argmax_{\parm \in \RR^\dimparm} \sum_{i = 1}^N w_i(\rx) \ell_i(\parm).
\end{eqnarray}
The log-likelihood contribution $\ell_i$ of the $i$th subject is obtained
from the unconditional model
\begin{eqnarray} \label{fm:trafo}
\Prob(\rY \le \ry) = 1 - \exp\left(-\exp\left(\basisy(\ry)^\top \parm
\right)\right). 
\end{eqnarray}
Ignoring possible truncation, we obtain the following contributions to the log-likelihood
\citep{Hothorn_Moest_Buehlmann_2016}:
\begin{eqnarray*}
\ell_i(\parm) = \left\{\begin{array}{ll}
\basisy(\ry_i)^\top \parm - \exp\left(\basisy(\ry_i)^\top \parm\right) + 
  \log(\basisy^\prime(\ry_i)^\top \parm) & \rY = \ry_i \\
-\exp\left(\basisy(\ubar{\ry}_i)^\top \parm \right) & \rY > \ubar{\ry}_i \\
\log\left(1 -\exp\left(-\exp\left(\basisy(\bar{\ry}_i)^\top \parm \right)\right)\right) & \rY < \bar{\ry}_i \\
\log\left(\exp\left(-\exp\left(\basisy(\ubar{\ry}_i)^\top \parm \right)\right) -
  \exp\left(-\exp\left(\basisy(\bar{\ry}_i)^\top \parm\right)\right)\right) & \rY \in
(\ubar{\ry}_i, \bar{\ry}_i].
\end{array} \right. 
\end{eqnarray*}
For an exact survival time $\ry_i$, $\basisy^\prime$ are the derivatives of
the basis functions $\basisy$.  In (\ref{fm:hatparm}), nearest-neighbor
weights $w_i(\rx)$ are obtained from a survival forest.  The weight
$w_i(\rx)$ is large, and thus the $i$th observation influences
$\hat{\parm}_N(\rx)$ when $\rx$ is similar to $\rx_i$.  Roughly speaking, this
similarity is measured by the number of times $\rx$ and $\rx_i$ end up in
the same terminal node of the trees constituting the forest.  The weight is
close to zero when $\rx$ and $\rx_i$ are elements of distinct terminal nodes
for most trees in the forest.  This aggregation scheme has been around for
some time \citep{Hothorn_Lausen_Benner_2004,Meinshausen_2006,Lin_Jeon_2006}
but only recently led to a more general understanding of random forests
\citep{Athey_Tibshirani_Wager_2018,Hothorn_Zeileis_2017}.  Although it seems
that the nearest-neighbor weights $w_i$, and thus the underlying survival
forest, are not linked to the log-likelihood function $\ell_i$ in
(\ref{fm:hatparm}), good performance can be achieved by implementing split
statistics that are sensitive to changes in the model parameters $\parm$
\citep{Athey_Tibshirani_Wager_2018,Hothorn_Zeileis_2017,Schlosser_Hothorn_Stauffer_2018}.

Forests of trees based on log-rank splitting (RF-S, CForest, Ranger) search
for splits by maximizing a two-sample log-rank test statistic over certain
binary splits in the prognostic variables.  The corresponding log-rank
scores are equivalent to the score contributions of an intercept $\alpha$ in
the model
\begin{eqnarray} \label{fm:Cox}
\Prob(\rY \le \ry) = 1 - \exp\left(-\exp\left(\basisy(\ry)^\top \parm + \alpha \right)\right), \quad \alpha = 0
\end{eqnarray}
with corresponding log-likelihood contributions $\ell_i(\parm, \alpha)$. The scores
\begin{eqnarray*}
s^{\alpha}_i(\parm) = \left.\frac{\partial \ell_i(\parm, \alpha)}{\partial \alpha}\right|_{\alpha = 0} \in \RR
\end{eqnarray*}
are called log-rank scores (technically, the term is used when only a
non-parametric form of the log-cumulative baseline hazard $\basisy(\ry)^\top
\parm$ is employed) and are powerful in detecting proportional hazards
deviations from model (\ref{fm:Cox}) of the form $\alpha(\rx) \neq \alpha =
0$.  The scores do not carry much information in a non-proportional hazards
setting, and thus the split statistic used in RF-S, CForest, or Ranger is
not very powerful in detecting potential splits in this situation.  Analytic
formulae for $s^{\alpha}_i$ and the more complex scores below have been
published elsewhere \citep{Hothorn_Moest_Buehlmann_2016}.

Based on model (\ref{fm:trafo}), we can construct split statistics that are
sensitive also in the non-proportional hazards setting, that is,
to deviations from the unconditional model of the form $\parm(\rx) \neq
\parm = \text{const}$. The corresponding scores are
\begin{eqnarray*}
\s_i(\parm) = \frac{\partial \ell_i(\parm)}{\partial \parm} \in \RR^\dimparm
\end{eqnarray*}
and appropriate test statistics are defined \citep{Hothorn_Zeileis_2017}. 
Trees based on these novel split statistics are now \emph{designed} to
detect changes in the conditional survivor function that are not well
described under the proportional hazards model~(\ref{fm:Cox}).

An additional advantage of the model-based view on survival forests
discussed here is the possibility of enriching models (\ref{fm:trafo}) or
(\ref{fm:Cox}) with additional parameters.  Predictive models feature an
additional treatment effect parameter $\eshiftparm$ that captures changes in
the conditional survivor function induced by a specific treatment ($\rz
= 0$ for placebo and $\rz = 1$ for Riluzole treatment in our case).  In the
simplest situation, the model
\begin{eqnarray*}
\Prob(\rY \le \ry \mid \rZ = \rz) = 1 - \exp\left(-\exp\left(\basisy(\ry)^\top \parm + \alpha + \eshiftparm \I(\rz = 1) \right)\right), \quad \alpha = 0
\end{eqnarray*}
leads to the bivariate score contributions
\begin{eqnarray*}
\s^{\alpha}_i(\parm, \eshiftparm) = \left.\frac{\partial \ell_i(\parm, \alpha, \eshiftparm)}{\partial (\alpha, \eshiftparm)^\top}\right|_{\alpha = 0} =
s_i^\alpha(\parm) (1, \I(\rz = 1)) \in \RR^2.
\end{eqnarray*}
Split statistics based on these scores have power against deviations of the
form $\alpha(\rx) \neq \alpha = 0$ and $\eshiftparm(\rx) \neq \eshiftparm =
\text{const}$, \ie in the proportional hazards setting.

Following the same reasoning as for prognostic models, we can relax the
proportional hazards assumption for the prognostic part, predictive part, or both
parts of the model. The model
\begin{eqnarray*}
\Prob(\rY \le \ry \mid \rZ = \rz) = 1 - \exp\left(-\exp\left(\basisy(\ry)^\top \parm + \eshiftparm \I(\rz = 1) \right)\right)
\end{eqnarray*}
allows non-proportional effects for the prognostic part $\parm(\rx)$
but still assumes differential treatment effects $\eshiftparm(\rx)$ as additive
effects on the scale of the log-cumulative hazard function. The
corresponding scores
\begin{eqnarray*}
\s_i(\parm, \eshiftparm) = \frac{\partial \ell_i(\parm, \eshiftparm)}{\partial (\parm^\top,
\eshiftparm)^\top}
\in \RR^{\dimparm + 1}
\end{eqnarray*}
can thus be used to define corresponding split statistics. If
non-proportional predictive effects are of special interest, the model
\begin{eqnarray} \label{fm:BsTrt}
\Prob(\rY \le \ry \mid \rZ = \rz) = 1 - \exp\left(-\exp\left(\basisy(\ry)^\top \parm +  \basisy(\ry)^\top
\parm_{\text{tr}}\I(\rz = 1) \right)\right)
\end{eqnarray}
defines time-varying (and thus non-proportional) differential 
treatment effects $\basisy(\ry)^\top \parm_\text{tr}(\rx)$ with scores
\begin{eqnarray*}
\s_i(\parm, \parm_{\text{tr}}) = \frac{\partial \ell_i(\parm, \parm_{\text{tr}})}{\partial
(\parm^\top, \parm^\top_{\text{tr}})^\top}
\in \RR^{2 \dimparm}.
\end{eqnarray*}

The primary aim of this study is to compare survival forests based on
log-rank scores to survival forests based on the novel general scores in the
prognostic and predictive settings.  Because a meaningful forest log-likelihood 
was defined for the survial setting herein, the performance of
survival forests can be evaluated by means of the out-of-sample log-likelihood defined by the
log-likelihood contributions $\ell_\imath$ of validation subjects $\imath =
1, \dots, \tilde{N}$:
\begin{eqnarray*}
\sum_{\imath = 1}^{\tilde{N}} \ell_\imath(\hat{\parm}_N(\rx_\imath)).
\end{eqnarray*}
This performance measure allows us to compare the impact of the choice of
the split statistic without taking into account the different aggregation
schemes used in different implementations of survival forests.  Only the
nearest neighbor weights $w_i$ are computed differently by the different
survival forest algorithms.  The same aggregation scheme (\ref{fm:hatparm})
based on the log-likelihood contributions $\ell_i$ obtained from
(\ref{fm:trafo}) for prognostic models and from (\ref{fm:BsTrt}) for
predictive models is used for all types of survival forests hereafter.  As
an additional feature of our model-based approach to survival forests, the
negative log-likelihood also defines a risk function for a novel
permutation-based variable importance applicable to the survival setting.

\begin{table}[t!]
\caption{Model Parameterizations. Overview of prognostic (without treatment
effect) and predictive (with treatment effect) models and corresponding
scores defining split statistics in survival trees. Each cell
contains a label for the combination of basis functions (columns) and
scores (rows). References to publications suggesting these models are given when applicable;
cells without citation correspond to novel developments. The
parameter $\alpha$ indicates a prognostic effect under proportional hazards,
and $\eshiftparm$ describes a predictive treatment effect in the same
situation. Under non-proportional hazards, prognostic effects are denoted by
$\parm$ and predictive (treatment) effects are denoted by $\parm_\text{tr}$. $\emptyset$ 
refers to computationally infeasible combinations. $\text{NP}(\alpha,
\eshiftparm)$ is conceptually computable but currently not implemented. 
\label{tab:models}}
\label{tab:methods}
\begin{center}
\begin{tabular}{p{.01\textwidth}l|p{.25\textwidth}cp{.25\textwidth}}
& & $\basisy_\text{W}$ & $\bern{\dimparm - 1}$ & $\basisy_\text{NP}$ \\ \hline
\multirow{2}{*}{\rotatebox[origin=r]{90}{Prognostic}}  & $s_i^{\alpha}(\parm)$ & $\text{W}(\alpha)$ & $\text{Bs}(\alpha)$ & $\text{NP}(\alpha)$ \newline \citep{Hothorn_Lausen_Benner_2004, Ishwaran_Kogalur_Blackstone_2008, Wright_Dankowski_Ziegler_2017} \\
& $\s_i(\parm)$ & $\text{W}(\parm)$ \newline \citep{Seibold_Zeileis_Hothorn_2017,Schlosser_Hothorn_Stauffer_2018} & $\text{Bs}(\parm)$ & $\emptyset$ \\ \hline 
 \multirow{3}{*}{\rotatebox[origin=r]{90}{Predictive}} & $\s_i^{\alpha}(\parm, \eshiftparm)$ & $\text{W}(\alpha,\eshiftparm)$ & $\text{Bs}(\alpha,\eshiftparm)$ & $\text{NP}(\alpha, \eshiftparm)$ \\
& $\s_i(\parm, \eshiftparm)$ & $\text{W}(\parm,\eshiftparm)$ \newline \citep{Seibold_Zeileis_Hothorn_2017}  & $\text{Bs}(\parm,\eshiftparm)$ & $ \emptyset$ \\
& $\s_i(\parm, \parm_\text{tr})$ & $\text{W}(\parm,\parm_\text{tr})$ & $\text{Bs}(\parm,\parm_\text{tr})$ & $\emptyset$ \\
\end{tabular}
\end{center}
\end{table}

In addition to different split statistics, different model parameterizations
have been suggested in the past.  RF-S, CForest, and Ranger are based on
non-parametric (NP) basis functions $\basisy_\text{NP}(\ry)^\top \parm$ that
assign one parameter to each observed event time.  In this case,
$\hat{\parm}$ is never explicitly computed; instead, the non-parametric
maximum-likelihood estimator for the unconditional survivor function
($\hat{S}_N$, for example, Kaplan-Meier or Breslow) is used:
$\basisy_\text{NP}(\ry)^\top \hat{\parm} = \text{cloglog}(1 -
\hat{S}_N(\ry))$, where cloglog is the complementary log-log function.  In a
parametric setting, Weibull (W)
models with basis functions $\basisy_\text{W}(\ry) = (1, \log(\ry))$ were
studied \citep{Seibold_Zeileis_Hothorn_2017}.  The
corresponding log-cumulative hazard function $\basisy(\ry)^\top \parm =
\eparm_1 + \eparm_2 \log(\ry)$ with $\parm = (\eparm_1, \eparm_2)^\top$
features one intercept parameter $\eparm_1$ and an accelerator $\eparm_2 >
0$.  As a compromise between the strict parametric setting and the
non-parametric setting, the application of Bernstein polynomials (Bs) has
been suggested \citep{McLain_Ghosh_2013,Hothorn_Moest_Buehlmann_2016}. 
Here, we suggest to use the basis functions $\basisy(\ry) = \bern{\dimparm -
1}(\log(\ry))$ (Bernstein polynomial of order $\dimparm - 1$ after
log-transformation) under the constraint that the log-cumulative hazard
function $\bern{\dimparm - 1}(\log(\ry))^\top \parm$ is non-decreasing
\citep[this constraint can be implemented as a linear constraint on the
parameters $\parm = (\eparm_1, \dots, \eparm_\dimparm)^\top \in
\RR^\dimparm$,][]{Hothorn_Moest_Buehlmann_2016}.  This choice allows simple
evaluation of the log-likelihood contributions $\ell_i$ while being
sufficiently flexible.  

An overview of the different models and their parameterizations is given in
Table~\ref{tab:models}.  We refer to forests using log-rank splitting
without specifying the baseline hazard function ($\text{NP}(\alpha)$, third
column in Table~\ref{tab:models}) as ``survival forests'', models based on a
conditional Weibull distribution (first column in Table~\ref{tab:models}) as
``distributional survival forests'' (DSF), and models based on a more
general parameterization via Bernstein polynomials (second column in
Table~\ref{tab:models}) as ``transformation survival forests'' (TSF).

The recently proposed $L_1$-splitting survival forests
\citep{Moradian_Larocque_Bellavance_2016} implement splits maximising the
integrated absolute difference between two survival functions, where the
corresponding groups are defined by a potential binary split.  The method
does not fit into the theoretical framework discussed here but was designed
to deal with non-proportional hazards and thus we compare it empirically to
the remaining forest variants in the next Section.

\section{Empirical Evaluation} \label{sec:empeval}

Survival forests (RF-S, CForest, Ranger), $L_1$-splitting survival forests,
distributional survival forests, and transformation survival forests were
evaluated empirically in both the prognostic and predictive setting assuming
a conditional Weibull data-generating process.  In the prognostic setting,
we were interested in a comparison of these random forest variants under
proportional hazards and under lack of proportionality of the hazard
functions in assessing the following hypotheses: (1) Weibull distributional
survival forests exactly matching the data-generating process outperform all
other methods.  (2) All methods perform similarly under proportional
hazards.  (3) Methods employing more general split statistics than log-rank
statistics ($L_1$, DSF W$(\parm)$, TSF Bs($\parm$)) perform better 
than log-rank-based forests (RF-S, CForest, Ranger, DSF W($\alpha$), TFS
Bs($\alpha$)) under non-proportional hazards.  Furthermore, we
were interested in quantifying the loss induced by using a too flexible
baseline hazard function (\eg two parameters in $\eparm_1 + \eparm_2
\log(t)$ versus $\dimparm = 6$ parameters in $\bern{\dimparm -
1}(\log(\ry))^\top \parm$) when comparing distributional survival forests to
transformation survival forests.  In the predictive setting, we compared the
two methods that are able to incorporate differential treatment effects,
namely distributional survival forests and transformation survival forests. 
In this case, we were interested in the loss associated with a too simple or
too complex choice of the model defining the split statistics.

\subsection{Weibull Data-generating Processes}

In both the prognostic and predictive setting, we simulated survival times
$\rY$ from a Weibull distribution with the conditional distribution function
\begin{eqnarray*}
\Prob(\rY \le \ry \mid \rX = \rx, \rZ = \rz) = 1 - \exp(-\exp(\xi(\rx, \rz) \log(\ry) + \alpha(\rx, \rz))),
\end{eqnarray*}
which features conditional parameter functions $\xi$ (scale term) and $\alpha$
(shift term) as functions of prognostic variables $\rx$ (in the prognostic
setting, here $\rz \equiv 0$) or predictive and prognostic variables $\rx$
under treatment $\rz$ (in the predictive setting).  For $\xi(\rx, \rz)
\equiv 1$ and $\alpha(\rx, \rz) \equiv 0$, we have $\rY \mid \rX = \rx \sim
\WD(1, 1)$.  When $\xi(\rx, \rz) \equiv 1$, the log-hazard ratio is
$\alpha(\rx, \rz)$.  Non-proportional hazards can be obtained when $\xi(\rx,
\rz) \neq 1$.  Because we were not interested in studying the impact of
potential censoring (all forests studied here are at least able to deal with
random right-censoring), we evaluated all models with respect to the
out-of-sample log-likelihood difference $\ell(\hat{\parm}_N) - \ell(\xi,
\alpha)$ of the out-of-sample log-likelihood
\begin{eqnarray*}
\ell(\hat{\parm}_N) & = & \sum_{\imath = 1}^{\tilde{N}}
\ell_\imath(\hat{\parm}_N(\rx_\imath), \hat{\parm}_{\text{tr}, N}(\rx_\imath)) \\
\end{eqnarray*}
and the log-likelihood evaluated at the true parameters $\xi$ and $\alpha$:
\begin{eqnarray*}
\ell(\xi, \alpha) & = & \sum_{\imath = 1}^{\tilde{N}} \xi(\rx_\imath, \rz_\imath) \log(\ry_\imath) + 
\alpha(\rx_\imath, \rz_\imath) - \exp\big(\xi(\rx_\imath, \rz_\imath) \log(\ry_\imath) + \alpha(\rx_\imath, \rz_\imath)\big) \\
& & \quad + \log\big(\xi(\rx_\imath, \rz_\imath)\big) + \log\big(t^{-1}_\imath\big)
\end{eqnarray*}
for $\tilde{N}$ validation samples $(\ry_\imath, \rx_\imath, \rz_\imath), \imath = 1, \dots, \tilde{N}$.

Shift and scale conditional parameter functions ($\alpha$ and $\xi$) were
modelled using the function \citep{Friedman_1991}
\begin{eqnarray*}
\text{F}(x_1, x_2, x_3, x_4, x_5) = 10\sin(\pi x_1x_2) + 20(x_3 - .5)^2 + 10x_4 +
5x_5.
\end{eqnarray*}
The output of $\text{F}$ was scaled to the
$[-1.5, 1.5]$ interval, denoted below as $\text{F}^\star$. This
choice restricted hazard ratios $\exp(\alpha(\rx, \rz)) =
\exp(\text{F}^\star)$ to values between $\exp(-1.5)$ and $\exp(1.5)$.

In the prognostic setting, we have $\alpha(\rx, \rz) = \alpha(\rx)$ 
and $\xi(\rx, \rz) = \xi(\rx)$.
Four types of effects were simulated for low- and high-dimensional data: No
effect (``No''; $\xi(\rx) \equiv 1$ and $\alpha(\rx) \equiv 0$), proportional
hazards effect (``PH''; $\xi(\rx) \equiv 1$ and $\alpha(\rx) \neq 0$),
non-proportional hazards effect (``Non-PH''; $\xi(\rx) \neq 1$ and $\alpha(\rx)
\equiv 0$), and a combination of PH and non-PH (``Combined''; $\xi(\rx) \neq
1$ and $\alpha(\rx) \neq 0$).  The effects were defined as follows:
\begin{center}
\begin{tabular}{lll} \hline
  & $\alpha(\rx)$ & $\xi(\rx)$ \\ \hline
\textbf{No} & $0$ & $1$ \\
\textbf{PH} & $\text{F}^\star(x_1, \dots, x_5)$ & $1$ \\
\textbf{Non-PH} & $0$ & $\exp\big(\text{F}^\star(x_6, \dots, x_{10})\big)$ \\
\textbf{Combined} & $\text{F}^\star(x_1, \dots, x_5)$ & $\exp\big(\text{F}^\star(x_6, \dots, x_{10})\big)$ \\ \hline
\end{tabular}
\end{center}
Low-dimensional prognostic variables were modelled with $J = 15$ independent
uniform variables ($10$ prognostic variables and $5$ additional noise
variables), \ie $\rX = (X_1, \dots X_{J})$, $X_j \sim U[0, 1]$,
$j=1,\dots,J$.  High-dimensional prognostic variables were modelled in the
same manner, but with $J = 60$ independent uniform variables ($10$
prognostic variables and $50$ additional noise variables).

For the predictive setting, we used the same effect function
$\text{F}^\star$ for prognostic and predictive effects, but with
non-overlapping prognostic variables $x_1, \dots, x_5, x_{11}, \dots,
x_{15}$ and predictive variables $x_6, \dots, x_{10}, x_{16}, \dots,
x_{20}$:
\begin{center}
\begin{tabular}{lp{5cm}p{5cm}} \hline
  & $\alpha(\rx, \rz)$ & $\xi(\rx, \rz)$ \\ \hline
\textbf{PH} & $\text{F}^\star(x_1, \dots, x_5) +$ \newline $\text{F}^\star(x_6, \dots, x_{10})\I\{\rz = 1\}$ & $1$\\[10pt]
\textbf{Non-PH} & $0$ & $\exp\big(\text{F}^\star(x_{11}, \dots, x_{15}) +$ \newline
$\quad \text{F}^\star(x_{16}, \dots, x_{20})\I\{\rz = 1\}\big)$\\[10pt]
\textbf{Combined} & $\text{F}^\star(x_1, \dots, x_5) +$ \quad \newline $\text{F}^\star(x_6, \dots, x_{10})\I\{\rz = 1\}$ & $\exp\big(\text{F}^\star(x_{11}, \dots, x_{15}) +$ \newline $\text{F}^\star(x_{16}, \dots, x_{20})\I\{\rz = 1\}\big)$ \\ \hline
\end{tabular}
\end{center}
Low-dimensional prognostic and predictive variables were based on $J = 25$
independent uniform variables ($20$ informative variables and $5$ noise
variables).  For the high-dimensional setting, we used $J = 70$ independent
uniform variables ($20$ informative variables and $50$ noise variables).

\subsection{Prognostic Models} \label{sec:ProgMod}

We compared the seven prognostic models from the prognostic part of
Table~\ref{tab:methods} and, in addition, survival forests based on $L_1$
splitting \citep{Moradian_Larocque_Bellavance_2016}.  For all competitors
except Ranger, a common set of parameters was specified: $250$ trees of
maximal depth $10$ and not less than $20$ observations in a terminal node. 
For low-dimensional data, all $J$ variables were used for splitting in a
non-terminal node (\ie bagging was applied), while for high-dimensional
data, only a random subset ($\sqrt{J}$) of the variables was considered. 
Large trees meeting these restrictions were grown without any form of early
stopping.  Furthermore, all forests except Ranger were grown based on the
same sub-samples of the original observations.  Transformation survival
forests TSF $\text{Bs}(\alpha)$ and TSF $\text{Bs}(\parm)$ applied Bernstein
basis functions of order five to log-transformed survival times.  The
current Ranger implementation does not allow sub-samples and a maximum tree
depth to be specified.  Therefore, we approximated the above parameter
settings by restricting the size of a terminal node to a number of
observations computed as the maximum of $20$ and
the size of the learning sample divided by $2^10$.  We
repeated each of the eight simulation scenarios (four effect types in low
and high dimensions) $100$ times with learning and validation samples of
size $N = 250$ and $\tilde{N} = 500$, respectively.

\begin{figure}
\centering
\begin{knitrout}
\definecolor{shadecolor}{rgb}{0.969, 0.969, 0.969}\color{fgcolor}
\includegraphics[width=\maxwidth]{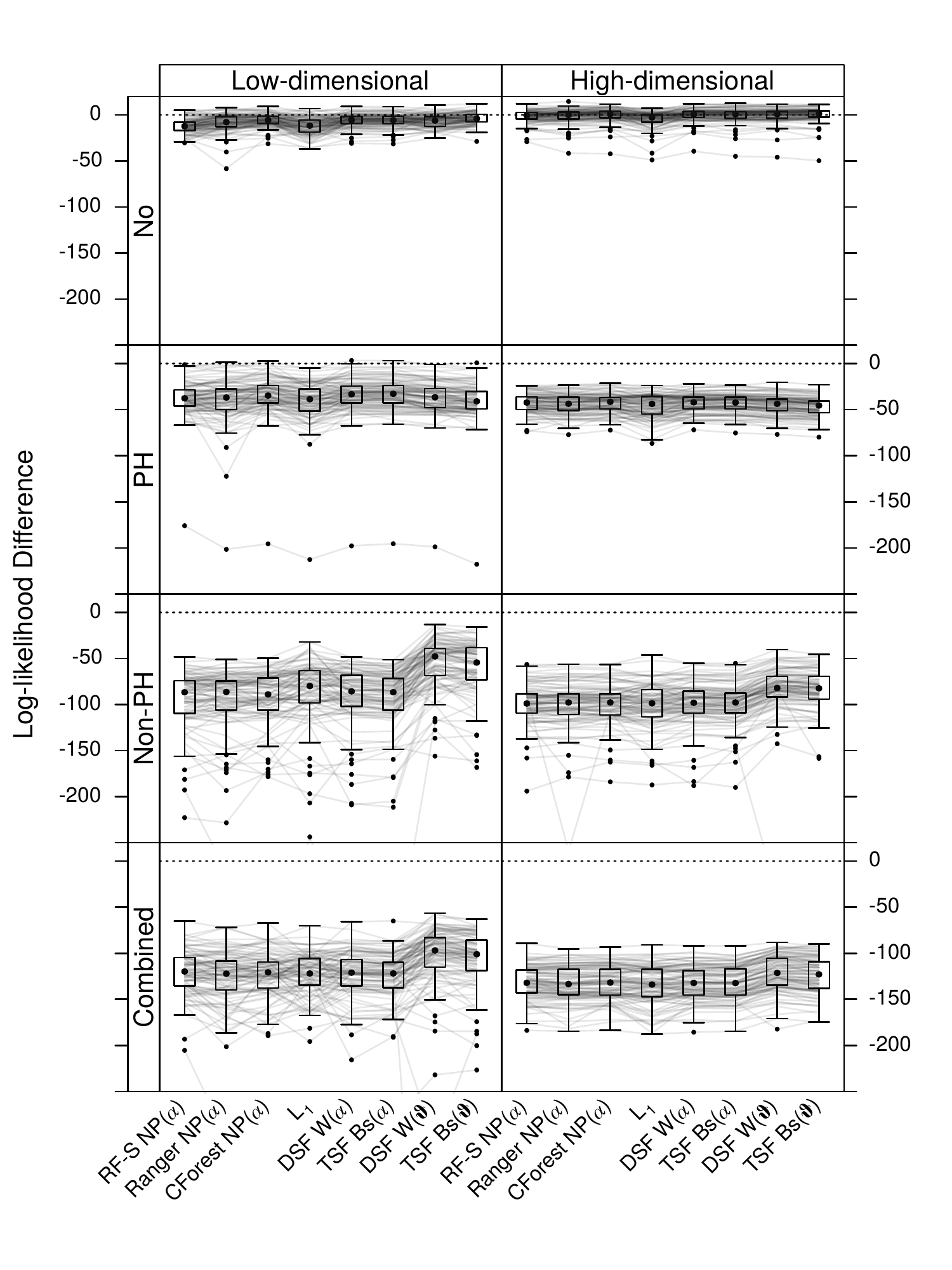} 

\end{knitrout}
\caption{Prognostic data-generating process. Performance of competitors
(random survival forest, RF-S; Ranger; conditional inference forests, CForest; $L_1$ survival forests;
transformation survival forests, TSF; and distributional survival forests, DSF; see Table~\ref{tab:models}) 
assessed by the difference between the out-of-sample
log-likelihood of the competitor and the log-likelihood of the true generating process.
Larger values of the difference are preferable. Eight scenarios, \ie low- and high-
dimensional prognostic variables for absent (No), proportional hazards
(PH), non-proportional hazards (Non-PH), or a combination of PH and Non-PH
(Combined) effects with $100$ repetitions each were simulated. Values
smaller than $-250$ are not shown.}
\label{fig:res_prog}
\end{figure}

The distribution of the out-of-sample log-likelihood differences
$\ell(\hat{\parm}_N) - \ell(\xi, \alpha)$ are presented in
Figure~\ref{fig:res_prog}.  In the absence of any effect (first row of
Figure~\ref{fig:res_prog}), roughly the same degree of overfitting was
observed for all competitors except RF-S and $L_1$.  The latter two
procedures exhibited a more pronounced overfitting.  In the presence of a
sole proportional hazards effect (second row of Figure~\ref{fig:res_prog}),
all competitors showed roughly the same performance.  Regardless of whether
the classical log-rank scores (based on the non-parametric basis functions
$\basisy_\text{NP}$) or scores obtained from (\ref{fm:Cox}) featuring
log-linear (DSF) or Bernstein (TSF) basis functions were applied, the
log-likelihood difference did not seem to be affected.  The loss induced by
a too rich parameterization ($\text{W}(\alpha)$ vs.~$\text{W}(\parm)$ and
$\text{Bs}(\alpha)$ vs.~$\text{Bs}(\parm)$) was negligible.  In the
non-proportional hazard setting (third row of Figure~\ref{fig:res_prog}),
the distributional survival forest splitting in both the scale and shift
parameters $\parm = (\eparm_1, \eparm_2)^\top$ clearly outperformed all competitors except for
the transformation survival forests that split in $\parm \in \RR^6$.  As
expected, $L_1$ forests were also able to pick-up this non-proportional
signal, but to a lesser degree.  All procedures employing log-rank splitting
assuming proportional hazards performed similarly.  The same conclusions
could be drawn for the combined proportional and non-proportional effects
setting, but the performance boost induced by the novel split criteria was
less pronounced.  The presence of $50$ variables in the high-dimensional
setting only marginally affected the performance of all methods tested.

\subsection{Predictive Models} \label{sec:PredMod}

\begin{figure}[t!]
\centering
\begin{knitrout}
\definecolor{shadecolor}{rgb}{0.969, 0.969, 0.969}\color{fgcolor}
\includegraphics[width=\maxwidth]{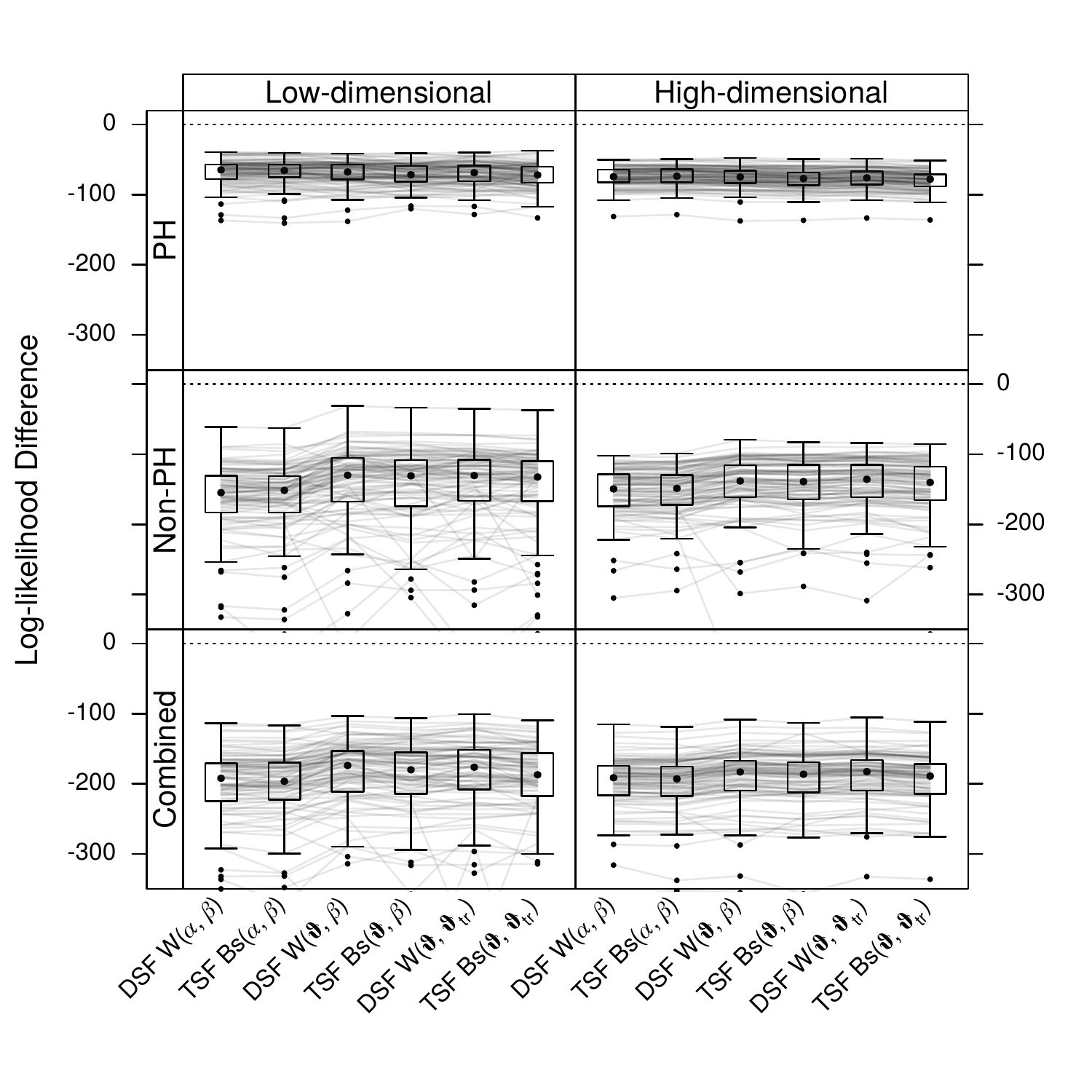} 

\end{knitrout}
\caption{Predictive data-generating process. Performance of the predictive
competitors distributional survival forests (DSF) and transformation survival
forests (TSF) in the presence of a simulated predictive effect was measured as the
difference between the out-of-sample log-likelihood of the competitor
and the true generating process.  Larger values of the difference are
preferable.  Six scenarios, i.e. low and high dimensions and proportional hazards (PH),
non-proportional hazards (Non-PH) and a combination of PH
and non-PH effects (Combined) with $100$ repetitions were
simulated.
\label{fig:res_pred}} 
\end{figure}

We compared the performance of the novel transformation survival trees in
the presence of a predictive effect to the performance of Weibull
distributional survival trees \citep{Seibold_Zeileis_Hothorn_2017}, \ie the six predictive models from the
first two columns of Table~\ref{tab:methods} were compared.  The same
parameter settings as described in Section~\ref{sec:ProgMod} were applied. 
In addition, subsamples were stratified with respect to treatment
assignment.  We again compared the out-of-sample log-likelihood difference
in the six scenarios (three effect types in low and high dimensions) for
$100$ learning and validation samples of size $N = 250$ and $\tilde{N} =
500$, respectively (Figure \ref{fig:res_pred}).

As expected, we found no differential performance in the proportional
hazards setting (first row of Figure \ref{fig:res_pred}).  Forests employing
a split criterion sensitive to non-proportional effects performed better in
the presence of a non-constant scale effect $\xi$ (second row of Figure
\ref{fig:res_pred}).  To a somewhat lesser extent, the same effect was
observed when both proportional and non-proportional prognostic and
predictive effects were simulated (third row of Figure \ref{fig:res_pred}). 
The impact of additional noise variables was only marginal.

\section{Amyotrophic Lateral Sclerosis Survival}

The Pooled Resource Open-Access ALS Clinical Trials (PRO-ACT,
\url{https://nctu.partners.org/ProACT}) database contains longitudinal data
of ALS patients who participated in one of $16$ phase II and III trials and
one observational study.  This project was initiated by the non-profit
organization Prize4Life (\url{http://www.prize4life.org/}) to increase
knowledge about ALS \citep{Kueffner_Zach_Norel_2014}.  The database contains
information on a broad variety of patient characteristics, such as vital
signs, the patient's and family's history, and treatment information. 
Identification criteria, such as study centers, are not included in the
database.  From the PRO-ACT database, we generated a data set of
$N = 3306$ observations containing
survival time and censoring information as well as baseline patient
characteristics.  Because not all procedures are able to deal with missing
values in prognostic or predictive variables, a complete case analysis was
performed. A more detailed description of the final data set of
$N = 2711$ observations and $18$ patient characteristics is
available elsewhere \citep{Seibold_Zeileis_Hothorn_2017}.  To estimate the performance of the
different procedures on the data set, we generated $100$ random splits of
the data into learning and validation samples in a $3:1$ proportion, keeping
the proportions of treated patients and the proportion of patients with
right-censored overall survival time in all learning and validation samples
the same as in the initial data set.

All survival forest variants discussed in this paper were applied using 
the same hyper-parameter settings as for the simulation study, including 
the use of bagging (i.e.\ no random variable selection). The number of 
randomly selected variables for splitting was set equal to the square 
root of the total number of variables.
In addition to the out-of-sample log-likelihood of these competitors, the
out-of-sample log-likelihoods of the following linear Weibull and Cox models is
reported: 
\begin{compactitem}
\item Cox$()$: an unconditional Cox model that ignores patient characteristics,
\item Weibull$(\alpha)$: A prognostic Weibull model with proportional hazard 
    $\exp(\rx^\top \alphavec)$,
\item Cox$(\alpha)$: a prognostic Cox model with proportional hazard $\exp(\rx^\top
\alphavec)$,
\item Weibull$(\alpha, \beta)$: a predictive Weibull model with proportional hazards \\ $\exp(\rx^\top \alphavec +
\eshiftparm \I\{\rz = 1\} + \rx^\top \shiftparm \I\{\rz = 1\}$), \ie,
including all treatment inter\-actions,
\item Cox$(\alpha, \beta)$: a predictive Cox model with proportional hazards \\
$\exp(\rx^\top \alphavec +
\eshiftparm \I\{\rz = 1\} + \rx^\top \shiftparm \I\{\rz = 1\}$),
\end{compactitem}
where $\alphavec$ denotes linear prognostic effects of $\rx$ and $\shiftparm$ denotes linear
differential predictive effects of $\rx$.

\begin{figure}[t!]
\centering
\begin{knitrout}
\definecolor{shadecolor}{rgb}{0.969, 0.969, 0.969}\color{fgcolor}
\includegraphics[width=\maxwidth]{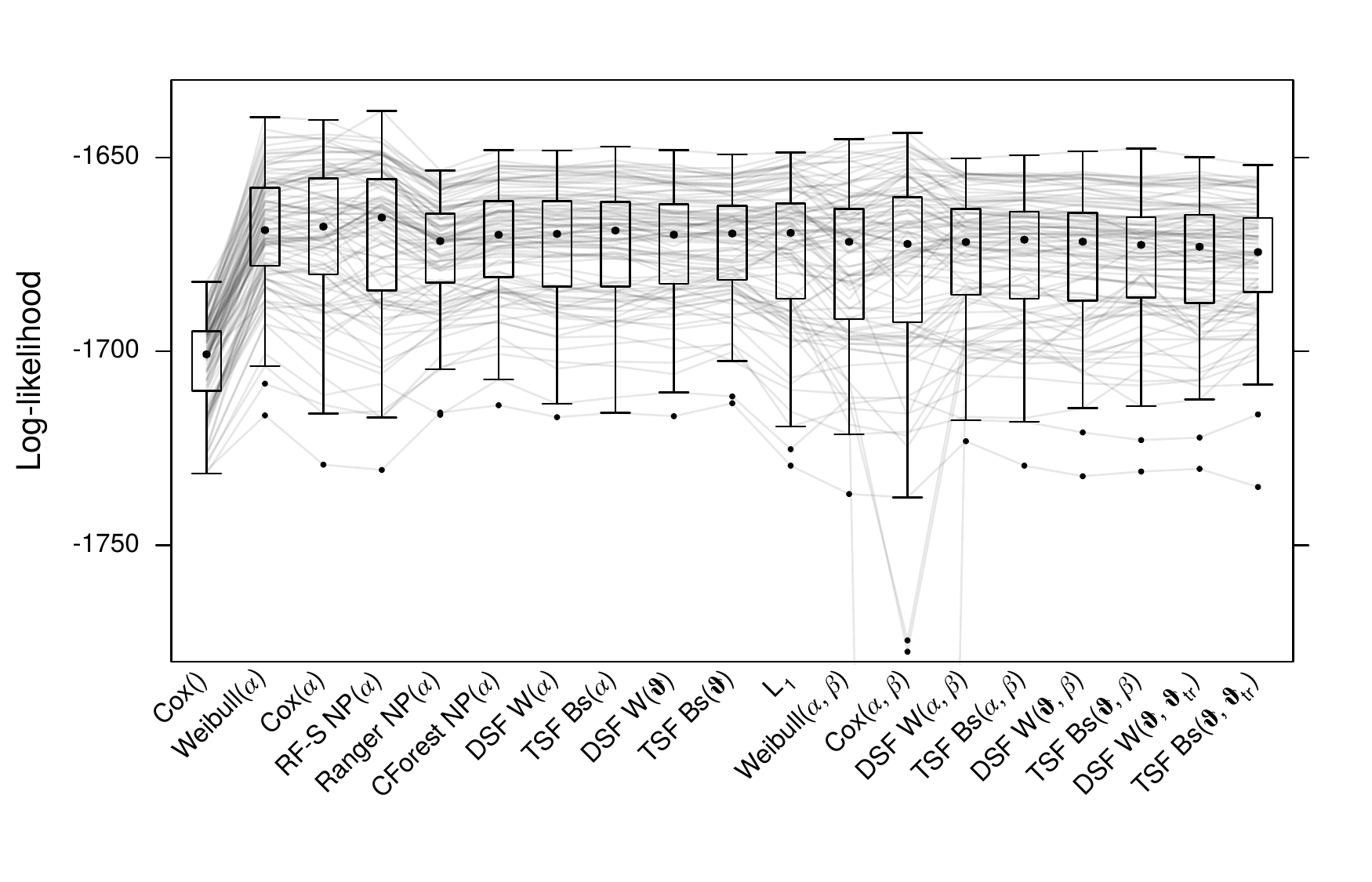} 

\end{knitrout}
\caption{ALS survival.
Out-of-sample log-likelihood estimation of prognostic and predictive models for $100$ random splits of ALS data into 
learning and validation samples with preserved treatment-censorship 
proportions. Larger values are preferable. One extreme value (for
Cox$(\alpha, \beta)$) is not shown.}
\label{fig:res_ALS}
\end{figure}

All methods taking patient characteristics into account outperformed the
unconditional Cox model (Figure~\ref{fig:res_ALS}).  We conclude that both
prognostic and predictive models gain their superior performance by
extracting information on patient's survival time from the corresponding
patient characteristics.  Among the prognostic competitors, the linear
Weibull and Cox models performed better than any of the survival forests
except RF-S.  Differences were, however, only marginal.  This is a strong
indication that neither non-linear interaction effects nor non-proportional
hazard effects were necessary to capture the signal in the data.  It is
worth noting that different parameterizations of distributional and
transformation survival forests performed highly similarly.

Predictive models did not noticeably better perform than prognostic models,
which confirms that the treatment effect is very weak.  Linear Weibull and
Cox models that included treatment-covariate interactions performed as well
as any of the distributional or transformation forests.  Again, variants of
the latter two procedures had only minor differences.

\begin{table}[t!]
\small
\begin{center}
\begin{tabular}{ll|l|ll}
 & & Prognostic & \multicolumn{2}{c}{Predictive} \\
Variable & Category & $\exp(\alphavec)$ & $\exp(\alphavec)$ & $\exp(\eshiftparm), \exp(\shiftparm)$ \\ \hline
Treatment  &  Riluzole  &     &    &  1.39 (0.04, 43.58) \\
Time since onset  &    &  1.37 (1.28, 1.47)  &  1.24 (1.10, 1.39)  &  1.10 (0.95, 1.28) \\
Race  &  Asian  &  1.78 (0.56, 5.62)  &  1.52 (0.41, 5.61)  &  1.30 (0.13, 12.58) \\
  &  African A.  &  2.48 (1.02, 6.04)  &  3.10 (0.98, 9.83)  &  1.46 (0.12, 17.72) \\
  &  Unknown  &  0.77 (0.47, 1.26)  &  1.65 (0.70, 3.90)  &  0.27 (0.09, 0.79) \\
Sex  &  Male  &  0.89 (0.74, 1.07)  &  1.14 (0.86, 1.51)  &  0.72 (0.49, 1.05) \\
Age (in yrs)  &    &  0.96 (0.95, 0.96)  &  0.95 (0.94, 0.96)  &  1.00 (0.99, 1.02) \\
Height (in cm)  &    &  1.01 (1.00, 1.02)  &  1.01 (1.00, 1.03)  &  1.00 (0.98, 1.02) \\
Atrophy  &  Yes  &  0.79 (0.50, 1.25)  &  1.29 (0.59, 2.84)  &  0.57 (0.21, 1.56) \\
Cramps  &  Yes  &  0.52 (0.36, 0.75)  &  0.61 (0.33, 1.13)  &  0.81 (0.37, 1.75) \\
Fasciculations  &  Yes  &  1.09 (0.68, 1.75)  &  1.15 (0.54, 2.41)  &  1.11 (0.40, 3.06) \\
Gait changes  &  Yes  &  0.86 (0.45, 1.65)  &  4.23 (0.57, 31.25)  &  0.14 (0.02, 1.18) \\
Other changes  &  Yes  &  1.22 (0.70, 2.12)  &  1.47 (0.64, 3.41)  &  0.96 (0.29, 3.18) \\
Sensory changes  &  Yes  &  1.27 (0.65, 2.49)  &  0.92 (0.39, 2.16)  &  1.75 (0.46, 6.66) \\
Speech  &  Yes  &  0.73 (0.58, 0.90)  &  0.87 (0.61, 1.23)  &  0.72 (0.46, 1.14) \\
Stiffness  &  Yes  &  1.56 (0.77, 3.15)  &  2.11 (0.64, 6.94)  &  0.62 (0.14, 2.73) \\
Swallowing  &  Yes  &  0.97 (0.63, 1.51)  &  1.21 (0.54, 2.72)  &  0.97 (0.35, 2.67) \\
Weakness  &  Yes  &  0.69 (0.58, 0.82)  &  0.73 (0.53, 1.00)  &  0.90 (0.61, 1.33) \\
Family (Older)  &  Yes  &  1.05 (0.86, 1.27)  &  0.93 (0.69, 1.26)  &  1.12 (0.76, 1.65) \\
Family (Same)  &  Yes  &  0.93 (0.70, 1.23)  &  0.98 (0.63, 1.53)  &  0.90 (0.51, 1.60) \\
Family (Younger)  &  Yes  &  1.30 (0.64, 2.62)  &  1.31 (0.47, 3.69)  &  0.99 (0.24, 4.14) \\

\end{tabular}
\caption{ALS survival. Estimated hazard ratios and corresponding unadjusted $95\%$
confidence intervals from prognostic and predictive linear
Weibull models learned on the ALS data ($N = 2711$ observations,
$851$ of whom died). The reference category for treatment
is placebo; that for race is Caucasian. Time
since onset was measured in years, and family history was coded as three
binary variables (older relatives affected by ALS, relatives in the same or
younger generation).}
\label{tab:ALS_Weib}
\end{center}
\end{table}

The results of this model evaluation indicate that simple linear prognostic
or predictive Weibull models can be used to adequately describe the impact
of patient characteristics on the survival time.  We estimated hazard ratios
with unadjusted $95\%$ confidence intervals of prognostic and predictive
Weibull models, based on the entire ALS data set of $N = 2711$
complete cases (Table~\ref{tab:ALS_Weib}).  In the prognostic model, seven
variables strongly affected the outcome (time since onset, race, age,
height, cramps, speech, and weakness).  In the predictive model, only three
prognostic variables (time since onset, age, weakness) in the presence of
one predictive contrast (unknown race) affected the outcome.  The
permutation variable importance, using the log-likelihood of the
corresponding trees as error function, of the prognostic distributional
survival forest W$(\alpha)$ and the predictive distributional survival
forests W$(\alpha, \beta)$ qualitatively coincided with the findings of
the linear Weibull models, \ie the variables time since onset, age, height,
cramps, speech, and weakness were more important than the remaining
variables in the prognostic model.  The variables time since onset, age, and
weakness showed up in the predictive distributional survival forest.

\begin{figure}[t!]
\begin{center}
\begin{knitrout}
\definecolor{shadecolor}{rgb}{0.969, 0.969, 0.969}\color{fgcolor}
\includegraphics[width=\maxwidth]{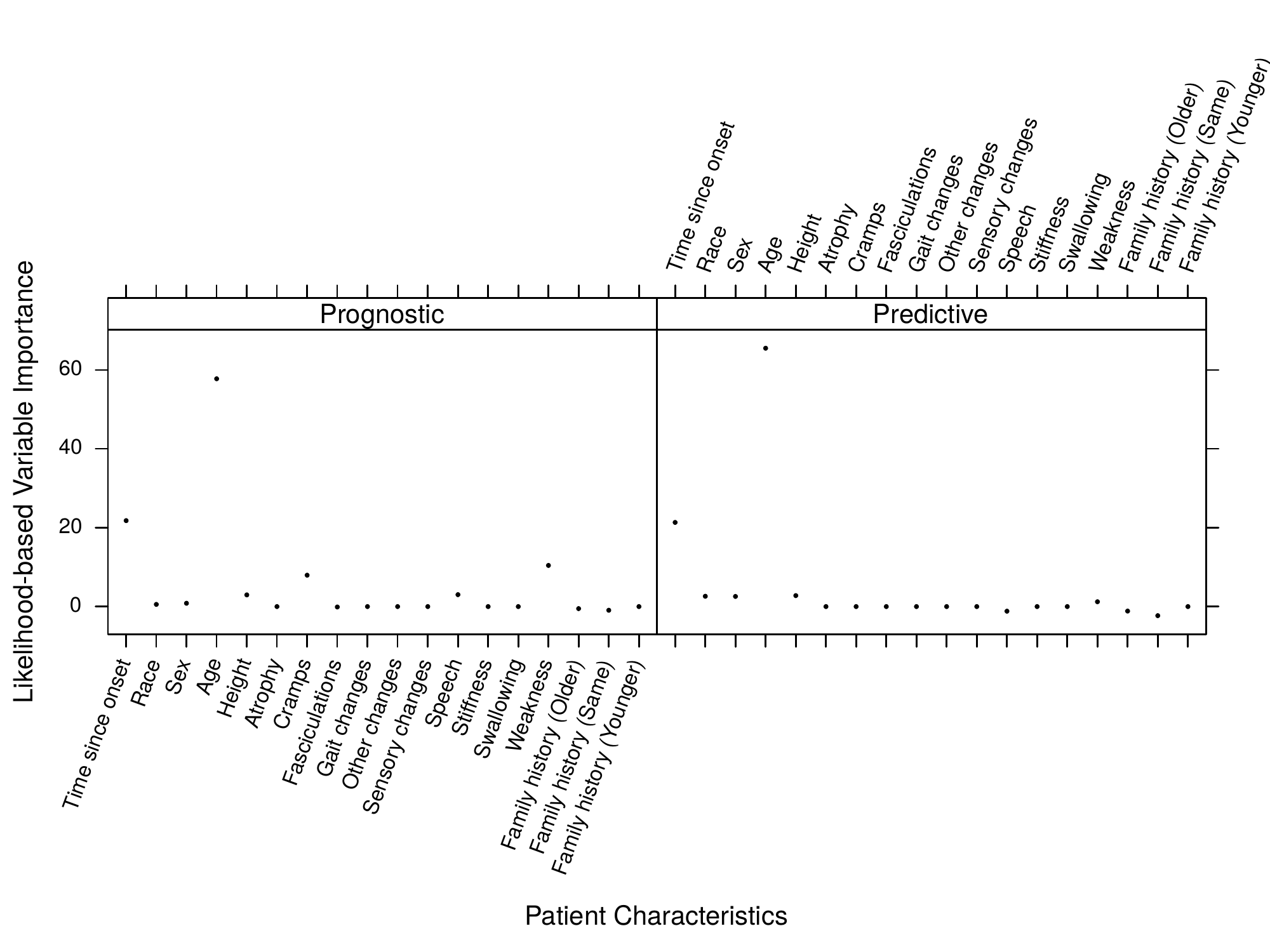} 

\end{knitrout}
\caption{ALS survival. Likelihood-based permutation variable importance of prognostic (left) and
  predictive (right) distributional survival forests DSF W$(\alpha)$ and DFS
  W$(\alpha, \beta)$, respectively.}
\end{center}
\end{figure}

\section{Discussion}

The authors of the first regression tree method \citep{MorganSonquist1963},
called automated interaction detection (AID), motivated the need for such a
procedure to overcome the limitations of linearity and additivity in linear
regression.  In the same spirit, modern successors of AID are commonly
understood as representatives of non-parametric regression methods.  With
the rise of statistical and machine learning, the superb accuracy of, for
example, random forests \citep{Breiman_2001} with its poor interpretability
on the one hand, and the often poor accuracy but excellent interpretability
of classical linear models on the other hand, motivated the dichotomous
understanding of algorithmic versus parametric modelling cultures
\citep{Breiman_2001_SS}.

While assumptions like additivity and linearity could, in fact, be
successfully overcome in the algorithmic modelling culture, other classical
assumptions inherent in the parametric modelling culture did not likewise
magically disappear. It was earlier demonstrated
\citep{Athey_Tibshirani_Wager_2018, Hothorn_Zeileis_2017} that random forests rely on
homogeneous residual variances and, consequently, quantile regression
forests \citep{Meinshausen_2006} are unable to adapt to patterns where only
the variance depends on certain explanatory variables.  Here, we used a
similar line of argumentation to demonstrate that survival forests, or at
least prominent implementations that rely on trees based on log-rank split
statistics for cut-point estimation, inherit the assumption of proportional
hazards from the corresponding Cox model that defines the associated
log-rank score statistics.

From a parametric modelling point of view, model-based transformation
survival forests are fruitful in two ways.  First, the underlying Cox models
can be extended to allow time-varying effects.  Thus, patterns emerging
under non-proportional hazards can be described and, consequently, detected
by appropriate score statistics in survival trees and forests.  Second, it
is possible to enrich simple prognostic models with treatment effects such
that survival trees and forests for the identification of differential
treatment effects can be developed for randomized clinical trial data.

From a practical point of view, our re-analysis of the PRO-ACT database of
ALS patients demonstrated that neither non-linear, interaction, nor
non-proportional hazards effects are necessary to describe prognostic and
predictive models for ALS survival time.  Simple linear Weibull models
performed similarly to the most flexible transformation survival forests
introduced here.  Consequently, we gain simplicity of model interpretation
without compromising model accuracy.  Of course, this finding is mostly due
to a low signal-to-noise ratio in this specific long-standing and difficult
to address problem.

As a by-product, the novel distributional and transformation survival
forests are able to deal with random left-censoring and interval-censoring
as well as left-, right-, and interval truncation \citep[the necessary
changes to the likelihood and score functions are explained in][and are
implemented in the \pkg{trtf} package, see next
Section]{Hothorn_Moest_Buehlmann_2016}.  Thus, survival forests featuring
time-varying prognostic variables can be set up using these procedures.  The survival
forests discussed here extend currently proposed survival tree methods for
interval-censored data
\citep{Fu_Simonoff_2017,Drouin_Hocking_Laviolette_2017}.  The former method
relies on score statistics from a Cox model and thus inherits specific power
for detecting proportional-hazard-type signals.  The latter maximum margin
interval trees employ a specific Hinge loss adapted to the interval-censored
case.  The connection of this approach to proportional hazards models
remains to be investigated.  As an additional feature, the log-likelihood
function associated with distributional and transformation survival forests
allows permutation variable importance measures to be obtained also in the
presence of random censoring and truncation, thus waiving the need for
falling back on general scoring rules, such as the inverse probability of
censoring-weighted Brier score \citep{Graf_Schmoor_Sauerbrei_1999}.

A limitation of our study is the lack of attention paid to the impact caused
by the implementation of different aggregation schemes.  Because we were
exclusively interested in a fair comparison of different split statistics,
the same aggregation via local adaptive maximum-likelihood estimation was
applied to all types of survival forests studied herein.  However, RF-S,
Ranger, and $L_1$ survival forests aggregate by averaging on the cumulative
hazard scale whereas CForest computes nearest-neighbor weighted Kaplan-Meier
curves.  Future research shall focus on this additional and important
difference that distinguishes the wide range of survival forests available
to practitioners.

\section*{Computational Details}

All computations were performed using
\textsf{R} version 3.5.2
\citep{R}. The code for data preprocessing of the PRO-ACT data is available in the
\pkg{TH.data} add-on package \citep{pkg:TH.data}. Patient-level data are
available to registered users from \url{https://nctu.partners.org/ProACT}.
Distributional and transformation survival forests were computed
using the \code{traforest()} function from the \pkg{trtf} add-on package
\citep{pkg:trtf}.  Random survival forests were obtained from the 
\pkg{randomForestSRC} add-on package \citep{pkg:randomForestSRC}. The other
two survival forests based on log-rank split statistics were CForest
\citep[function \code{cforest()} from the \pkg{party} add-on package,][]{pkg:party}
and Ranger \citep[package \pkg{ranger},][]{pkg:ranger}. 
$L_1$ survival forests were computed with a privately patched
version of \pkg{randomForestsSRC} provided to the authors by Professor Denis Laroque,
HEC Montr\'eal, Canada.

The \pkg{trtf} package was built on top of the infrastructure packages
\pkg{partykit} \citep{Hothorn_Zeileis_2015} and \pkg{mlt}
\citep{Hothorn_2018_JSS}.  For the empirical evaluation in
Section~\ref{sec:empeval}, all survival forests except Ranger were fitted
using the same $250$ subsamples of size $.632 N$ (\pkg{ranger} version
0.11.1 does not allow subsamples to be
specified ).  Trees were restricted to at least $20$ observations in each
terminal node and a maximal tree depth of $10$.  None of the tree growing
algorithms applied internal prepruning.  For the low-dimensional
simulations, bagging was applied.  In all other settings and the analysis of
the ALS data, a random subset of size $\sqrt{J}$ of the available prognostic
or predictive variables was considered for splitting only (\code{mtry}
parameter).  For transformation survival forests, transformation functions
were parameterized in terms of Bernstein polynomials for log-time of order
five.  Log-likelihoods were optimized under monotonicity constraints using a
combination of augmented Lagrangian minimization and spectral projected
gradients.

For the curious reader, we provide a small example of how the
prognostic transformation survival forest TSF Bs$(\parm)$ and the predictive
transformation survival forest TSF Bs($\parm, \eshiftparm)$ can be estimated
for the publically available German Breast Cancer Study Group-2 data:
\small{
\begin{knitrout}
\definecolor{shadecolor}{rgb}{0.969, 0.969, 0.969}\color{fgcolor}\begin{kframe}
\begin{alltt}
\hlcom{### attach data and packages}
\hlkwd{data}\hlstd{(}\hlstr{"GBSG2"}\hlstd{,} \hlkwc{package} \hlstd{=} \hlstr{"TH.data"}\hlstd{)}
\hlkwd{library}\hlstd{(}\hlstr{"survival"}\hlstd{)} \hlcom{# CRAN: survival infrastructure}
\hlkwd{library}\hlstd{(}\hlstr{"tram"}\hlstd{)}     \hlcom{# CRAN: transformation models}
\hlkwd{library}\hlstd{(}\hlstr{"trtf"}\hlstd{)}     \hlcom{# CRAN: transformation trees and forests}
\hlkwd{set.seed}\hlstd{(}\hlnum{290875}\hlstd{)}    \hlcom{# make results reproducible}

\hlcom{### prognostic model for GBSG2}
\hlcom{## fit unconditional Cox model, with in-sample log-likelihood}
\hlkwd{logLik}\hlstd{(m_prog} \hlkwb{<-} \hlkwd{Coxph}\hlstd{(}\hlkwd{Surv}\hlstd{(time, cens)} \hlopt{~} \hlnum{1}\hlstd{,}
                       \hlkwc{data} \hlstd{= GBSG2,} \hlkwc{log_first} \hlstd{=} \hlnum{TRUE}\hlstd{))}
\end{alltt}
\begin{verbatim}
## 'log Lik.' -2638.152 (df=7)
\end{verbatim}
\begin{alltt}
\hlcom{## fit TSF(theta)}
\hlstd{TSF_prog} \hlkwb{<-} \hlkwd{traforest}\hlstd{(m_prog,} \hlkwc{formula} \hlstd{=} \hlkwd{Surv}\hlstd{(time, cens)} \hlopt{~} \hlstd{.,}
                      \hlkwc{data} \hlstd{= GBSG2)}
\hlcom{## compute out-of-bag log-likelihood}
\hlkwd{logLik}\hlstd{(TSF_prog,} \hlkwc{OOB} \hlstd{=} \hlnum{TRUE}\hlstd{)}
\end{alltt}
\begin{verbatim}
## 'log Lik.' -2596.698 (df=NA)
\end{verbatim}
\begin{alltt}
\hlcom{### predictive model for GBSG2}
\hlcom{## fit conditional Cox model with PH effect of hormonal treatment}
\hlkwd{logLik}\hlstd{(m_pred} \hlkwb{<-} \hlkwd{Coxph}\hlstd{(}\hlkwd{Surv}\hlstd{(time, cens)} \hlopt{~} \hlstd{horTh,}
                       \hlkwc{data} \hlstd{= GBSG2,} \hlkwc{log_first} \hlstd{=} \hlnum{TRUE}\hlstd{))}
\end{alltt}
\begin{verbatim}
## 'log Lik.' -2633.649 (df=8)
\end{verbatim}
\begin{alltt}
\hlcom{## fit TSF(theta, beta)}
\hlstd{TSF_pred} \hlkwb{<-} \hlkwd{traforest}\hlstd{(m_pred,}
                      \hlkwc{formula} \hlstd{=} \hlkwd{Surv}\hlstd{(time, cens)} \hlopt{|} \hlstd{horTh} \hlopt{~} \hlstd{.,}
                      \hlkwc{data} \hlstd{= GBSG2)}
\hlcom{## compute out-of-bag log-likelihood}
\hlkwd{logLik}\hlstd{(TSF_pred,} \hlkwc{OOB} \hlstd{=} \hlnum{TRUE}\hlstd{)}
\end{alltt}
\begin{verbatim}
## 'log Lik.' -2601.604 (df=NA)
\end{verbatim}
\end{kframe}
\end{knitrout}
}
Corresponding \code{predict()} methods allow computation of conditional
survivor or hazard functions as well as differential treatment effects
$\eshiftparm(\rx)$ from the resulting models.  Computing the distributional
survival forests only requires that the \code{Coxph()} function be replaced with a
call to \code{Survreg()}.  The code necessary to reproduce the empirical
results reported in this paper is available from within \textsf{R} 
\begin{knitrout}
\definecolor{shadecolor}{rgb}{0.969, 0.969, 0.969}\color{fgcolor}\begin{kframe}
\begin{alltt}
\hlkwd{system.file}\hlstd{(}\hlstr{"survival_forests"}\hlstd{,} \hlkwc{package} \hlstd{=} \hlstr{"trtf"}\hlstd{)}
\end{alltt}
\end{kframe}
\end{knitrout}

\bibliography{mlt,packages,ALS}



\end{document}